\documentclass[aps,twocolumn,showpacs,byrevtex,prl,reprint,nofootinbib]{revtex4-1}
\usepackage{xspace}

\usepackage{epsfig}
\usepackage{dcolumn}
\usepackage{bm}
\usepackage{color}
\usepackage{graphicx}
\usepackage{subfigure}
\usepackage{pstricks}
\usepackage{pst-node}
\usepackage{rotating}
\usepackage{times}
\usepackage[english]{babel}
\addto{\captionsenglish}{%

}
\usepackage[colorlinks,linkcolor=blue,citecolor=blue]{hyperref}
\usepackage{lineno}
\newcommand{\ssb}{\Sigma^0\bar{\Sigma}^0}

\newcommand{\XXN}{\Xi^{0}\bar\Xi^{0}}
\newcommand{\XXB}{\Xi^{-}\bar\Xi^{+}}
\newcommand{\XXXB}{\Xi(1530)^{-}\bar\Xi(1530)^{+}}
\newcommand{\XXBB}{\Xi(1530)^{-}\bar\Xi^{+}}

\newcommand{\EE}{e^+e^-}
\newcommand{\BB}{B\bar{B}}
\newcommand{\ppb}{p\bar{p}}
\newcommand{\psp}{\psi(3686)}
\newcommand{\jpsi}{J/\psi}
\newcommand{\ar}{\rightarrow}
\newcommand{\llb}{\Lambda\bar{\Lambda}}

\newcommand{\bfg}{\begin{figure}}
\newcommand{\efg}{\end{figure}}
\newcommand{\bitm}{\begin{itemize}}
\newcommand{\eitm}{\end{itemize}}
\newcommand{\bnum}{\begin{enumerate}}
\newcommand{\enum}{\end{enumerate}}
\newcommand{\btbl}{\begin{table*}}
\newcommand{\etbl}{\end{table*}}
\newcommand{\btbu}{\begin{tabular}}
\newcommand{\etbu}{\end{tabular}}
\newcommand{\bcl}{\begin{center}}
\newcommand{\ecl}{\end{center}}
\newcommand{\bbt}{\bibitem}
\newcommand{\beq}{\begin{equation}}
\newcommand{\eeq}{\end{equation}}
\newcommand{\beqr}{\begin{eqnarray}}
\newcommand{\eeqr}{\end{eqnarray}}

\begin{document}
\normalsize
\parskip=5pt plus 1pt minus 1pt
\title{\boldmath Observation of $\psp\ar\XXXB$ and $\XXBB$}
\author{
M.~Ablikim$^{1}$, M.~N.~Achasov$^{10,d}$, P.~Adlarson$^{59}$, S. ~Ahmed$^{15}$, M.~Albrecht$^{4}$, M.~Alekseev$^{58A,58C}$, A.~Amoroso$^{58A,58C}$, F.~F.~An$^{1}$, Q.~An$^{55,43}$, Y.~Bai$^{42}$
% ...
\\
\vspace{0.2cm}
(BESIII Collaboration)\\
\vspace{0.2cm} {\it
$^{1}$ Institute of High Energy Physics, Beijing 100049, People's Republic of China\\
% ...
$^{4}$ Bochum Ruhr-University, D-44780 Bochum, Germany\\
% ...
$^{10}$ G.I. Budker Institute of Nuclear Physics SB RAS (BINP), Novosibirsk 630090, Russia\\
% ...
$^{15}$ Helmholtz Institute Mainz, Johann-Joachim-Becher-Weg 45, D-55099 Mainz, Germany\\
% ...
$^{42}$ Southeast University, Nanjing 211100, People's Republic of China\\
$^{43}$ State Key Laboratory of Particle Detection and Electronics, Beijing 100049, Hefei 230026, People's Republic of China\\
% ...
$^{55}$ University of Science and Technology of China, Hefei 230026, People's Republic of China\\
% ...
$^{58}$ (A)University of Turin, I-10125, Turin, Italy; (B)University of Eastern Piedmont, I-15121, Alessandria, Italy; (C)INFN, I-10125, Turin, Italy\\
$^{59}$ Uppsala University, Box 516, SE-75120 Uppsala, Sweden\\
% ...
$^{d}$ Also at the Novosibirsk State University, Novosibirsk, 630090, Russia\\
% ...
}
}

\date{\today}

\begin{abstract}
 Using $448.1 \times 10^{6}$ $\psp$ events collected with the BESIII detector at BEPCII,
 we employ a single-baryon tagging technique to make the first observation of  $\psp\ar\XXXB$ and $\XXBB$ decays with a statistical significance of more than 10$\sigma$  and 5.0$\sigma$, respectively. The branching fractions are measured to be $\cal{B}$$(\psp\ar\XXXB) $ =  (11.45 $\pm$ 0.40 $\pm$ 0.59) $\times$ $10^{-5}$ and $\cal{B}$$(\psp\ar\XXBB)$ = (0.70 $\pm$ 0.11 $\pm$ 0.04) $\times$ $10^{-5}$.
The angular distribution parameter for $\psp\ar\XXXB$ is determined to be $\alpha$ = 0.40 $\pm$ 0.24 $\pm$ 0.06, which agrees with the theoretical predictions within 1$\sigma$.
The first uncertainties are statistical, and the second systematic.
\end{abstract}
\maketitle

The decays of the charmonium resonances, such as $\psp$, into baryon anti-baryon pairs ($\BB$) have been extensively studied as a useful test of perturbative quantum chromodynamics (QCD)~\cite{Farrar01, Farrar}. They proceed via the annihilation of the $c\bar{c}$ pair into three gluons for strong decays or a virtual photon for electromagnetic decays.
Within the context of SU(3) flavor symmetry, decays of charmonium to $\BB$ (i.e., $B_{1}\bar{B}_{1}$, $B_{8}\bar{B}_{8}$ and $B_{10}\bar{B}_{10}$, where $B_{1}$ is for baryon singlet, $B_{8}$ is for baryon octet and $B_{10}$ is for baryon decuplet) are allowed, but decays into octet-decuplet baryonic pairs ($B_{10}\bar{B}_{8}$) are forbidden~\cite{Asner:2008nq,KWetc}. However, many experimental results on $\jpsi\ar B_{10}\bar{B}_{8}$ decay~\cite{SU3B,Ablikim:2012uqz} indicate the presence of flavor SU(3) symmetry breaking. 
There is no previous experimental information on $\psp\ar B_{10}\bar{B}_{8}$, such as $\psp\ar\XXBB$ decay.

Due to hadron helicity conservation~\cite{Farrar01,Pais:1971wu}, the angular distributions for the process $\EE\ar\psp\ar\BB$ are given by 
\begin{equation}\label{alpha}
\frac{dN}{d\cos\theta_{B}}\propto1+\alpha\cos^{2}\theta_{B},
\end{equation}
where $\theta_{B}$ is the angle between one of the baryons and the $e^{+}$ beam direction in the $\EE$ center-of-mass (CM) system, and the $\alpha$ is the angular distribution parameter, which is widely investigated in theory and experiment~\cite{Franklin:1983ve,ppbref,angularSig}.
Many theoretical models, such as those considering quark mass effects~\cite{ppbref02}, and electromagnetic effects~\cite{ppbref01}, 
predict that the angular distribution parameter obeys $\alpha< 1$. 
The BES and BESIII collaborations measured the angular distribution of $\jpsi\ar\Sigma^0\bar\Sigma^0$, $\Sigma(1385)\bar\Sigma(1385)$ and obtained a negative 
$\alpha$ value, but with poor precision~\cite{angularSig, Ablikim:2016iym}. H.~Chen {\it et al.}~\cite{Chen:2006yn} noted 
that the angular distribution parameter for $\jpsi$ and $\psp\ar\BB$ could be negative when
re-scattering effects of $\BB$ in heavy quarkonium decays are taken into account.
Additional measurements of the $\alpha$ parameter are of interest to 
confront the various theoretical approaches. 
In other experiments, the angular distributions for charmonium decays to baryon pairs,
such as $\jpsi$ and $\psp\ar\ppb, \llb, \ssb$, $\Xi\bar\Xi$, $\Sigma(1385)\bar\Sigma(1385)$~\cite{ppbref, Ablikim:2006aw}, were reported.  Negative $\alpha$ values were found for the processes $\jpsi\ar\Sigma^0\bar\Sigma^0$, $\Sigma(1385)\bar\Sigma(1385)$, while for other processes $\alpha$ was either measured to be positive, or not measured. 
The BESIII experiment has a large data sample at the $\psp$ resonance, which can be used to verify the theoretical models for the process like $\psp\ar\XXXB$, for which the $\alpha$ value is predicted to be 0.18 and 0.31~\cite{ppbref02, ppbref01}.

In this paper, the observation of $\psp\ar\XXXB$ and $\XXBB$ decays is presented based on $448.1 \times 10^{6}$ $\psp$ events~\cite{Ablikim:2017wyh} collected with the BESIII detector at the BEPCII in 2009 and 2012.  Selection of $\psp\ar\XXXB$ and $\XXBB$ events via full reconstruction suffers from 
low efficiency.  To achieve a higher efficiency, we first reconstruct a $\Xi(1530)^-$, referred to as a $\Xi(1530)^{-}$ tag, and then search for a recoiling $\bar{\Xi}(1530)^+$ or $\bar\Xi^+$ signal (unless otherwise noted, charge conjugation ($c.c.$) is implied throughout the paper). 

To determine the detection efficiencies for $\psp\ar\XXXB$ and $\XXBB$,
100,000 simulated events are generated for each reconstructed mode.  
For the $\XXXB$ final state, the angular distribution is generated with 
the $\alpha$ value measured in this analysis, while for $\XXBB$ we 
use $\alpha = 1$ based on theory~\cite{Pang}.  
The $\Xi(1530)^-$ decays to the $\pi^{0(-)}\Xi^{-(0)}$ modes, 
with $\Xi^{0(-)}\to\pi^{0(-)}\Lambda$, $\Lambda\to p\pi^-$ and
$\pi^0\to\gamma\gamma$, are simulated using \textsc{evtgen}~\cite{evt2}, and 
the response of the BESIII detector is modeled with Monte Carlo (MC) 
simulations using a framework based on \textsc{geant}{\footnotesize 4}~\cite{geant4}.  
A detailed description of the BESIII detector is given in Ref.~\cite{BESIII}.
To study the potential backgrounds, an inclusive MC sample of 350 $\times$ $10^{6}$ $\psp$ decays is generated, where
the production of the $\psp$ resonance is simulated with the \textsc{kkmc} generator~\cite{kkmc},
the subsequent decays are processed via
\textsc{evtgen}~\cite{evt2} according to the measured branching fractions
provided by the Particle Data Group (PDG)~\cite{PDG2016}, and the
remaining unmeasured decay modes are generated with
\textsc{lundcharm}~\cite{lund}. 
Data collected at the CM energy of  3.65 GeV (off-peak
data sample, 44 pb$^{-1}$)~\cite{Ablikim:2017wyh} is used to estimate the
contamination from the continuum processes $\EE\ar\XXXB$ and $\XXBB$. 

Charged tracks are reconstructed in the main drift chamber (MDC) within an angular range of ($|\cos\theta|<0.93$, where $\theta$ is the polar angle with respect to the $e^{+}$ beam direction).  
Information on the specific energy deposition ($dE/dx$) in the MDC and from the time-of-flight (TOF) counters are combined to form particle identification (PID) confidence levels (CLs) for pion, kaon, and proton hypotheses. 
Each track is assigned to the particle type with the highest CL.  At least two negatively-charged pions and one proton are required.  
Photons are reconstructed from isolated showers in the electromagnetic calorimeter (EMC).  Energy deposited 
in the nearby TOF counters is included to improve the reconstruction
efficiency and energy resolution. Photon energies are required to be
greater than 25 MeV in the EMC barrel region ($|\cos\theta|<0.80$) or
greater than 50 MeV in the EMC end caps 
($0.86 < |\cos\theta| < 0.92$). Showers in between these angular regions 
are poorly reconstructed and are excluded. 
The EMC shower timing is required to be within the range [0, 700] ns, 
relative to the event start time, to suppress electronic noise and energy 
deposits unrelated to the analyzed event. 
The number of good photon candidates, $N_{\gamma}$, must satisfy $2 \leq N_{\gamma} \leq 15$  based on a simulated signal MC study. 

In order to reconstruct $\pi^{0}$ candidates, a one-constraint (1C) kinematic fit is applied to all $\gamma\gamma$ combinations, constraining the two-photon invariant mass to the nominal $\pi^0$ mass~\cite{PDG2016}. 
To suppress non-$\pi^{0}$ backgrounds, only combinations with $\chi^{2}_{1C} < 20$ are retained by optimizing the figure of merit FOM = $\frac{S}{\sqrt{S + B}}$, where $S$ is the number of signal events and $B$ is the number of background events, based on the MC simulation. 
The $\Lambda$ candidates are reconstructed from $p\pi^-$ pairs with an invariant mass within 5 MeV/$c^{2}$ of the nominal $\Lambda$ mass.  This interval is determined by optimizing FOM.
A secondary vertex fit~\cite{Xu:2009zzg} is performed on all $p\pi^{-}$ combinations; those with $\chi^{2} < 500$ are kept for further analysis.  To further suppress the background, the decay length of the $\Lambda$ is required to be positive.  In the case of multiple candidates, the one with an unconstrained mass closest to the nominal mass is retained as used in Refs.~\cite{Ablikim:2016iym}.  The $\Xi$ candidates are reconstructed by considering all $\pi\Lambda$ combinations within 10 MeV/$c^{2}$ of the nominal $\Xi$ mass.  
For $\Xi^{-}$ candidates, a secondary vertex-constrained fit is used, while for both charged and neutral $\Xi$, only the candidate closest to the nominal mass is retained when there is more than one per event. The decay length of the $\Xi^{-}$ is required to be positive to further suppress the backgrounds.
The $\Xi(1530)^{-}$ candidates are reconstructed in the $\pi^{0}\Xi^{-}$ and $\pi^{-}\Xi^{0}$ modes and the candidate closest to the nominal mass is retained when there is more than one per event.

\begin{figure}[!htbp]
\includegraphics[width=0.23\textwidth]{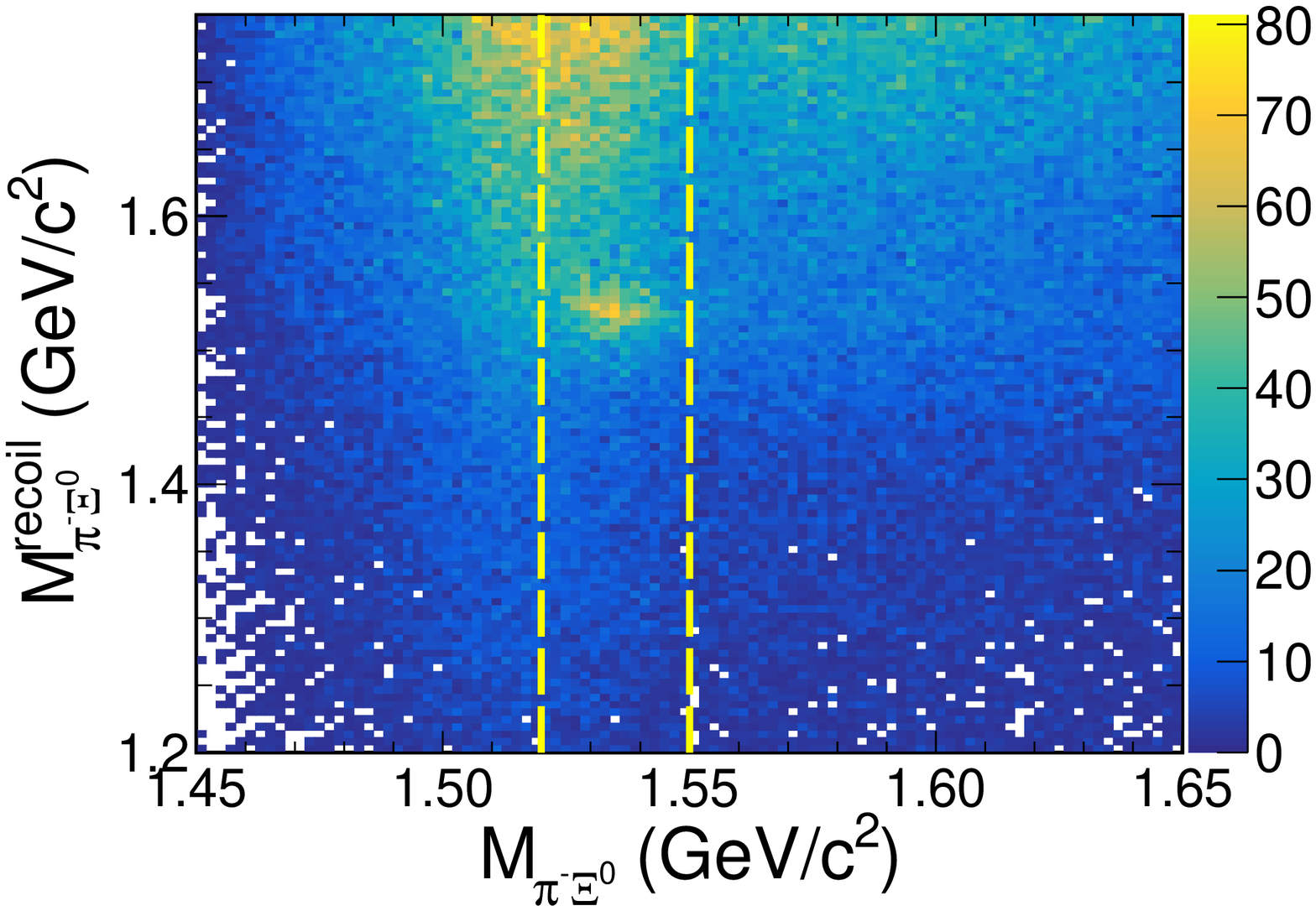}
\includegraphics[width=0.23\textwidth]{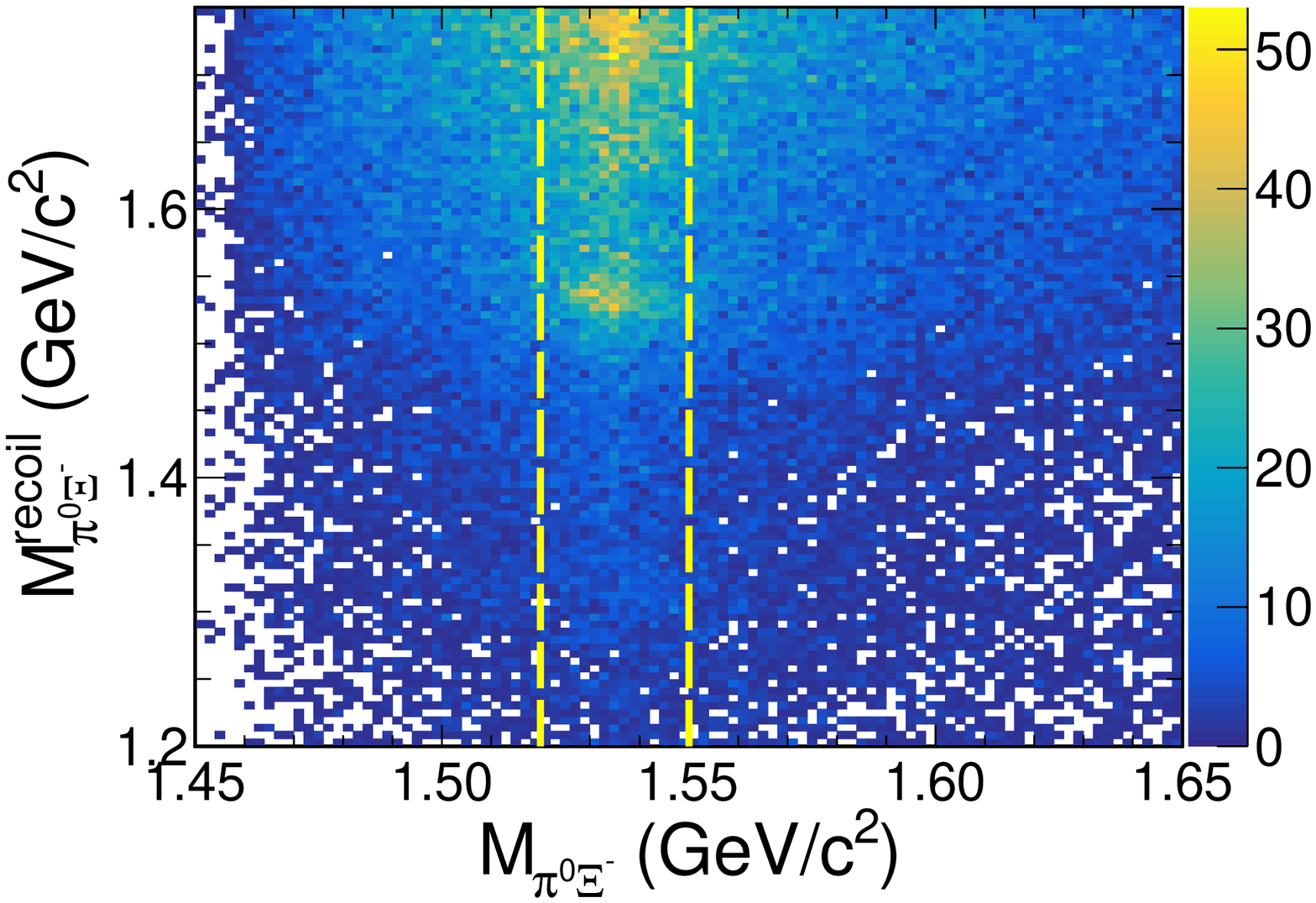}
\caption{Distributions of $M_{\pi\Xi}$ versus $M^{\rm recoil}_{\pi\Xi}$ for $\pi^{-}\Xi^{0}$ mode (Left) and $\pi^{0}\Xi^{-}$ mode (Right). The dashed lines denote the $\Xi(1530)$ signal region.}
\label{scatterplot}
\end{figure}

The anti-baryon candidates $\bar\Xi^{+}$ and $\bar\Xi(1530)^{+}$ are inferred by the mass recoiling against the selected $\pi\Xi$ system,
\begin{equation}
M^{\rm recoil}_{\pi\Xi} = \sqrt{(E_{\rm CM}-E_{\pi\Xi})^{2} - |\vec{p}_{\pi\Xi}|^{2}},
\end{equation}
where $E_{\pi\Xi}$ and $\vec{p}_{\pi\Xi}$ are the energy and momentum of the selected 
$\pi\Xi$ system, and $E_{\rm CM}$ is the CM energy.
Figure~\ref{scatterplot} shows the scatter plot of $M_{\pi\Xi}$ versus $M^{\rm recoil}_{\pi\Xi}$. 
To determine signal yields, the mass of the $\pi\Xi$ is required to be within 15 MeV/$c^{2}$ of the nominal mass of $\Xi(1530)^{-}$. 

Our inclusive MC sample reveals that the main background for $\psp\ar\XXXB$ and $\XXBB$ decays comes from $\psp\ar\pi^+\pi^-(\pi^0\pi^0)\jpsi$ with $\jpsi\ar\XXB$; it is 
distributed smoothly in the signal region of $M^{\rm recoil}_{\pi\Xi}$.   
Only a few events in the off-peak data sample survive and do not form any obvious peaking structures in the $\Xi(1530)$ signal region of the corresponding $M^{\rm recoil}_{\pi\Xi}$ distribution. Taking into account the normalization of the luminosity and CM energy dependence of the cross section, the contribution from continuum processes is expected to
be small and is neglected in the further analysis.
There are transition $\pi^0$s with similar momenta in both the baryon and anti-baryon decay chains within the signal events.  Incorrect use of these in the $\Xi^{0}$ or $\Xi(1530)^-$ reconstruction leads to a wrong combination background (WCB).  

\begin{figure*}[!htbp]
\bcl
\subfigure{\includegraphics[width=0.4\textwidth]{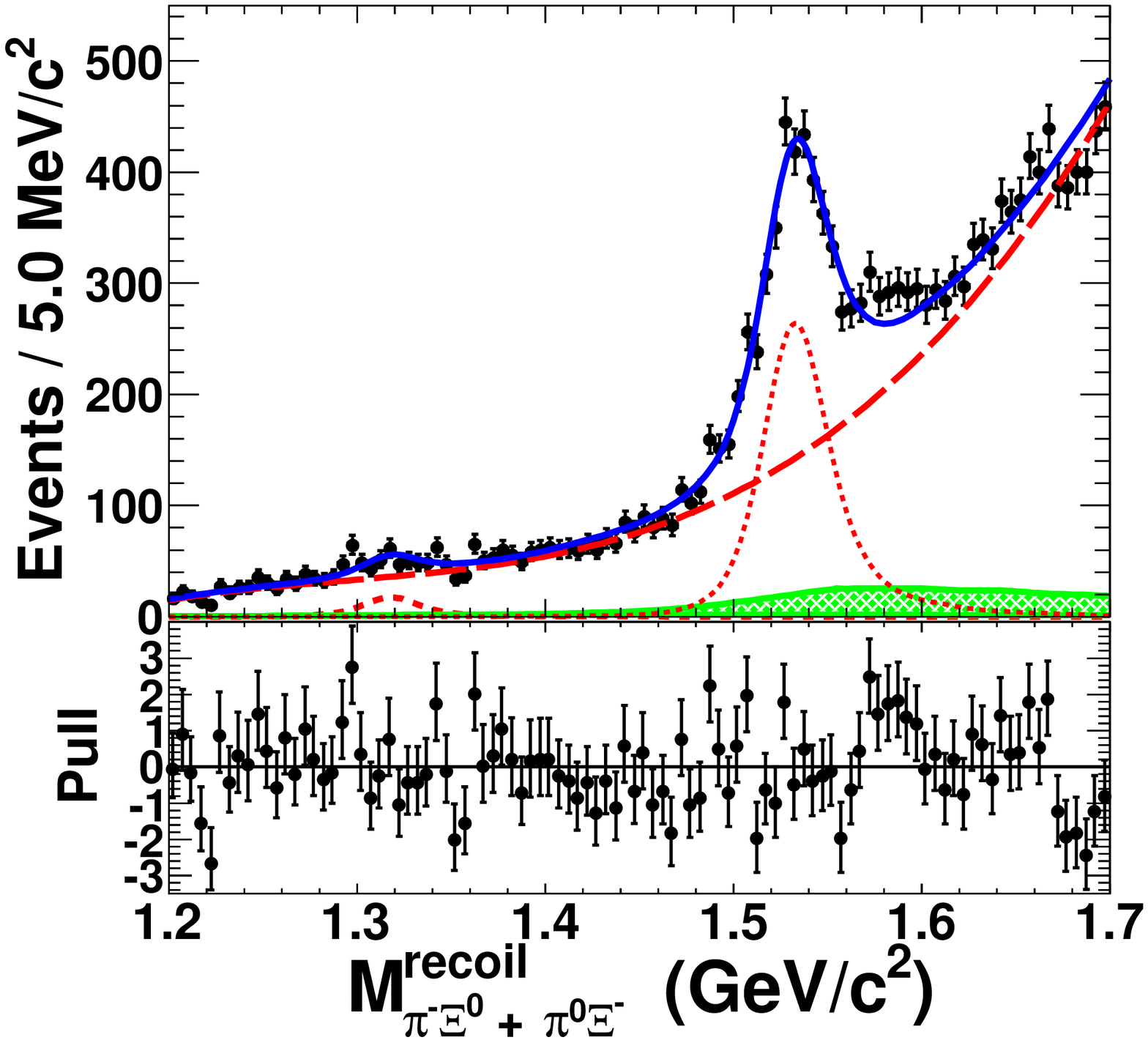}}
\put(-55.0,155){\Large (a)}
\put(-170.0,110){\subfigure{\includegraphics[width=0.16\textwidth]{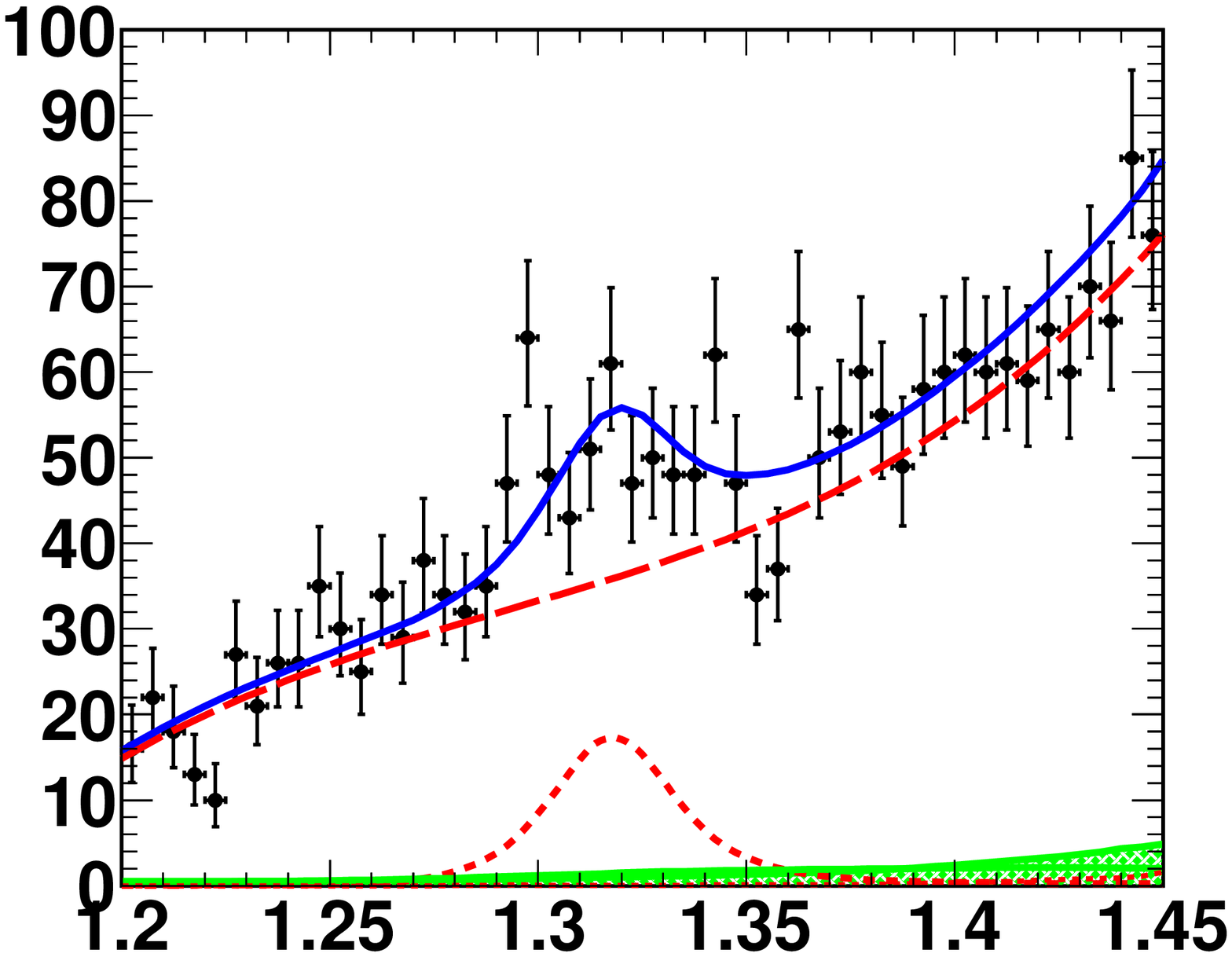}}}
\subfigure{\includegraphics[width=0.4\textwidth]{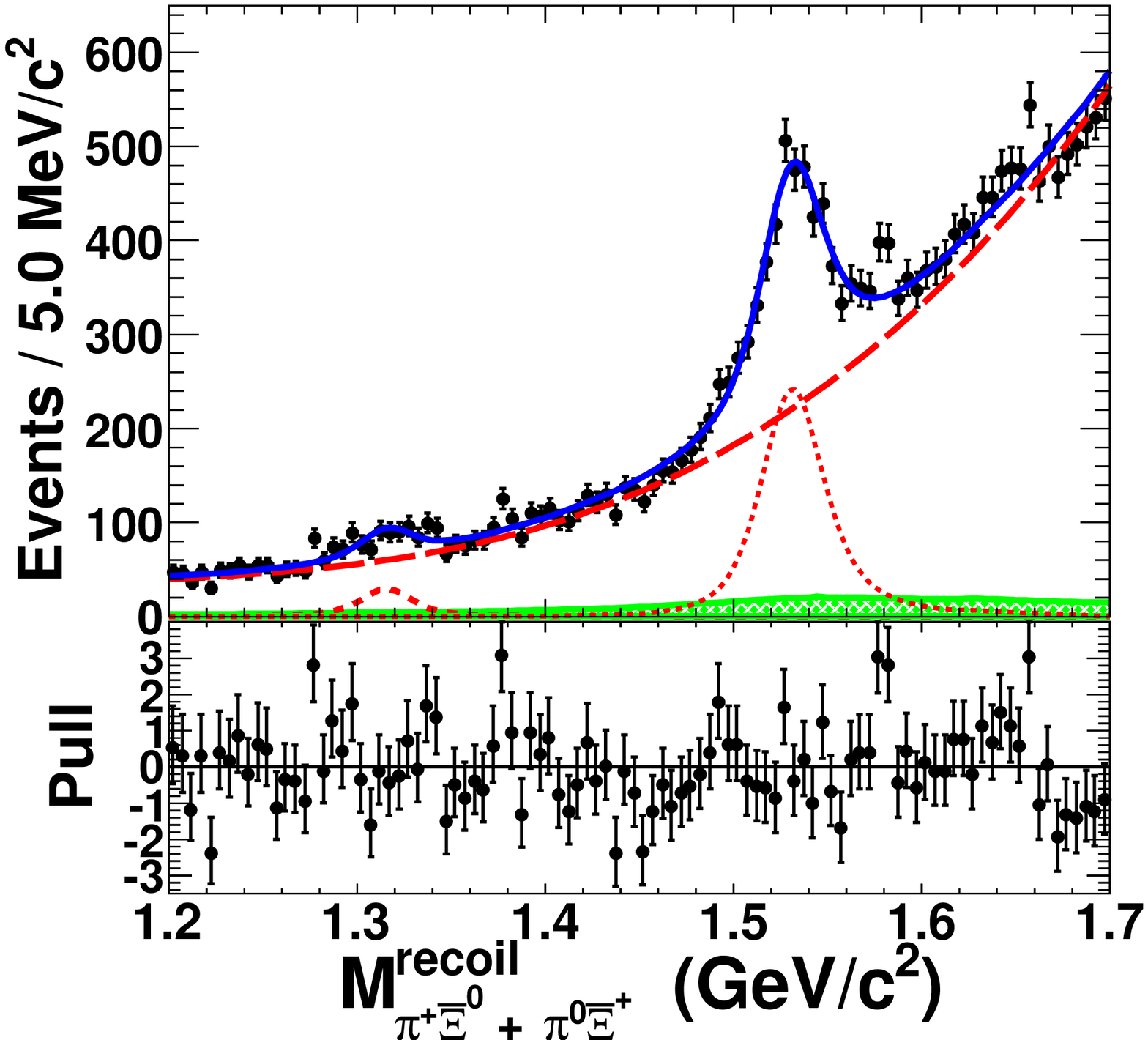}}
\put(-55.0,155){\Large (b)}
\put(-170.0,110){\subfigure{\includegraphics[width=0.16\textwidth]{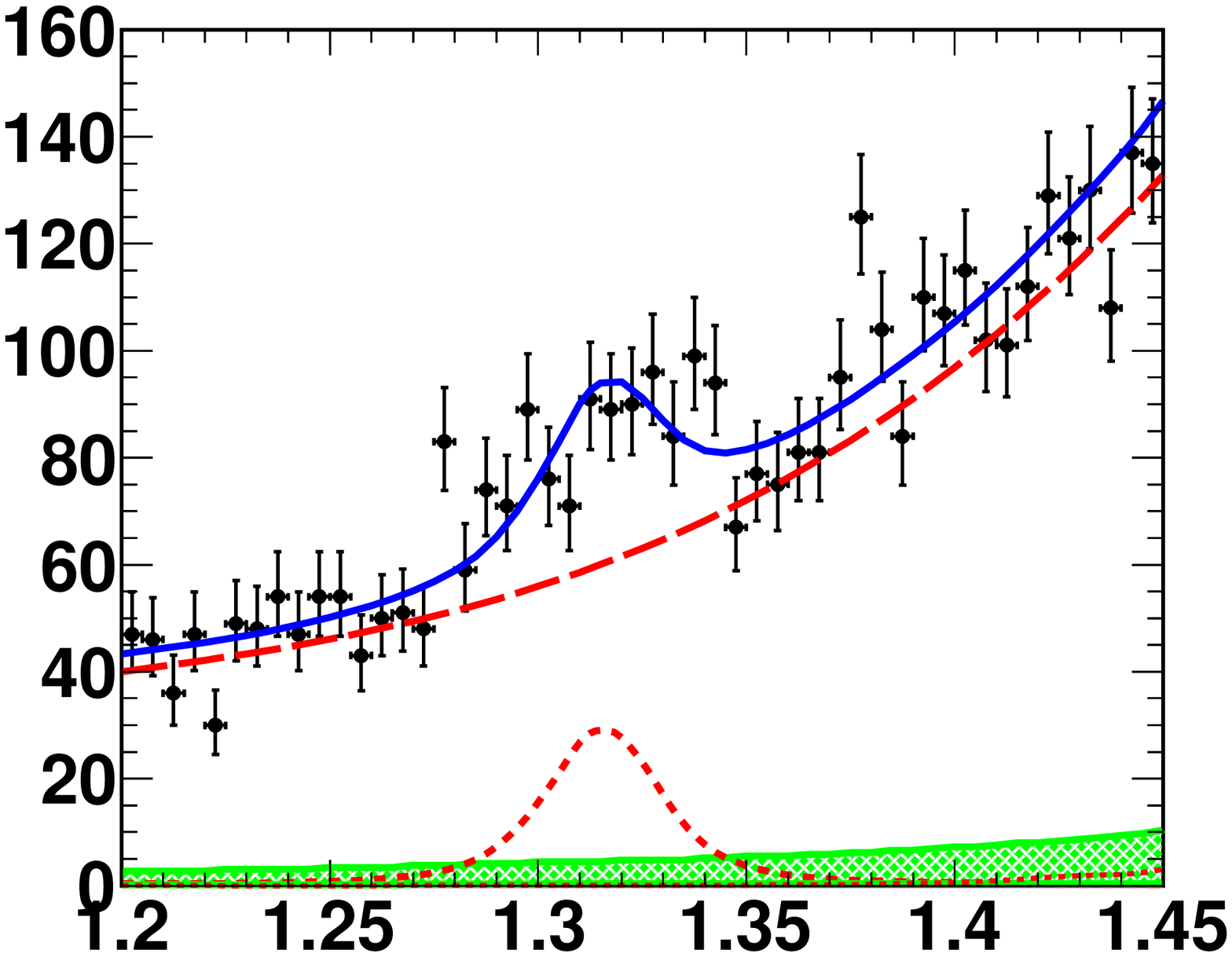}}}
\caption{
Fit to recoil mass spectra of $M^{\rm recoil}_{\pi^{-}\Xi^{0} + \pi^{0}\Xi^{-}}$ (a) and $M^{\rm recoil}_{\pi^{+}\bar\Xi^{0} + \pi^{0}\bar\Xi^{+}}$ (b).
Dots with error bars are for data, the blue solid lines show the fit result, the red short-dashed lines are for signal, the red long-dashed ones are for the smooth background (Other-Bkg), and the green hatched ones are for WCB. The  insets show the $\bar\Xi^+$ signal region in more detail.
}
\label{fitting}
\ecl
\end{figure*}

The signal yields for the two decays $\psp\ar\XXXB$ and $\XXBB$ are determined by performing an extended maximum likelihood fit to the $M^{\rm recoil}_{\pi\Xi}$ spectrum. In the fit, the signal shapes for the two decays are represented by the simulated MC shape convolved with a Gaussian function to take into account the mass resolution difference between the data and the MC simulation, 
where the parameters of the Gaussian function are left free but are shared by the two decay modes.  
The WCB is described by the simulated MC shape, and the corresponding numbers of events are fixed according to the MC simulation. The other remaining backgrounds (Other-Bkg) are found to distribute smoothly  in the $M_{\pi\Xi}^{\rm recoil}$ spectrum and are therefore described by a third-order Chebychev function.
Figure~\ref{fitting} shows the $M^{\rm recoil}_{\pi\Xi}$ distributions for the $\Xi(1530)^-$ and $\bar{\Xi}(1530)^+$ tags, respectively, with $\Xi$ and $\Xi(1530)$ peaks evident in each. Including systematic uncertainties, the significance for $\psp\ar\XXBB$ is calculated to be  more than $5.0 \sigma$ for the $\Xi(1530)^{-}\bar\Xi$ mode and its $c.c.$ mode combined.   The individual significances are calculated from the change in log likelihood and degrees of freedom with and without the signal in the fit.

\begin{figure}[!h]
\bcl
\includegraphics[width=0.22\textwidth]{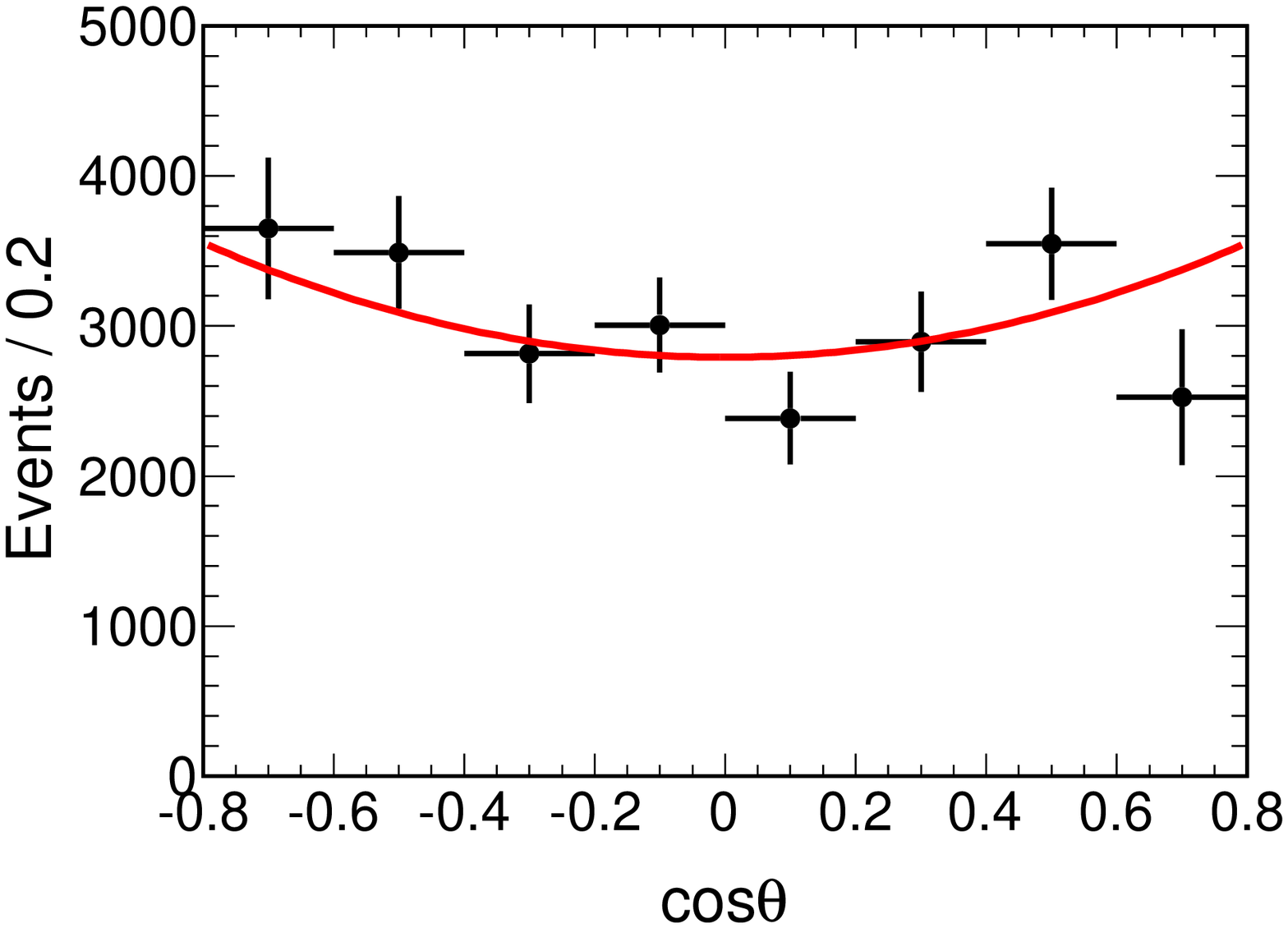}
\includegraphics[width=0.22\textwidth]{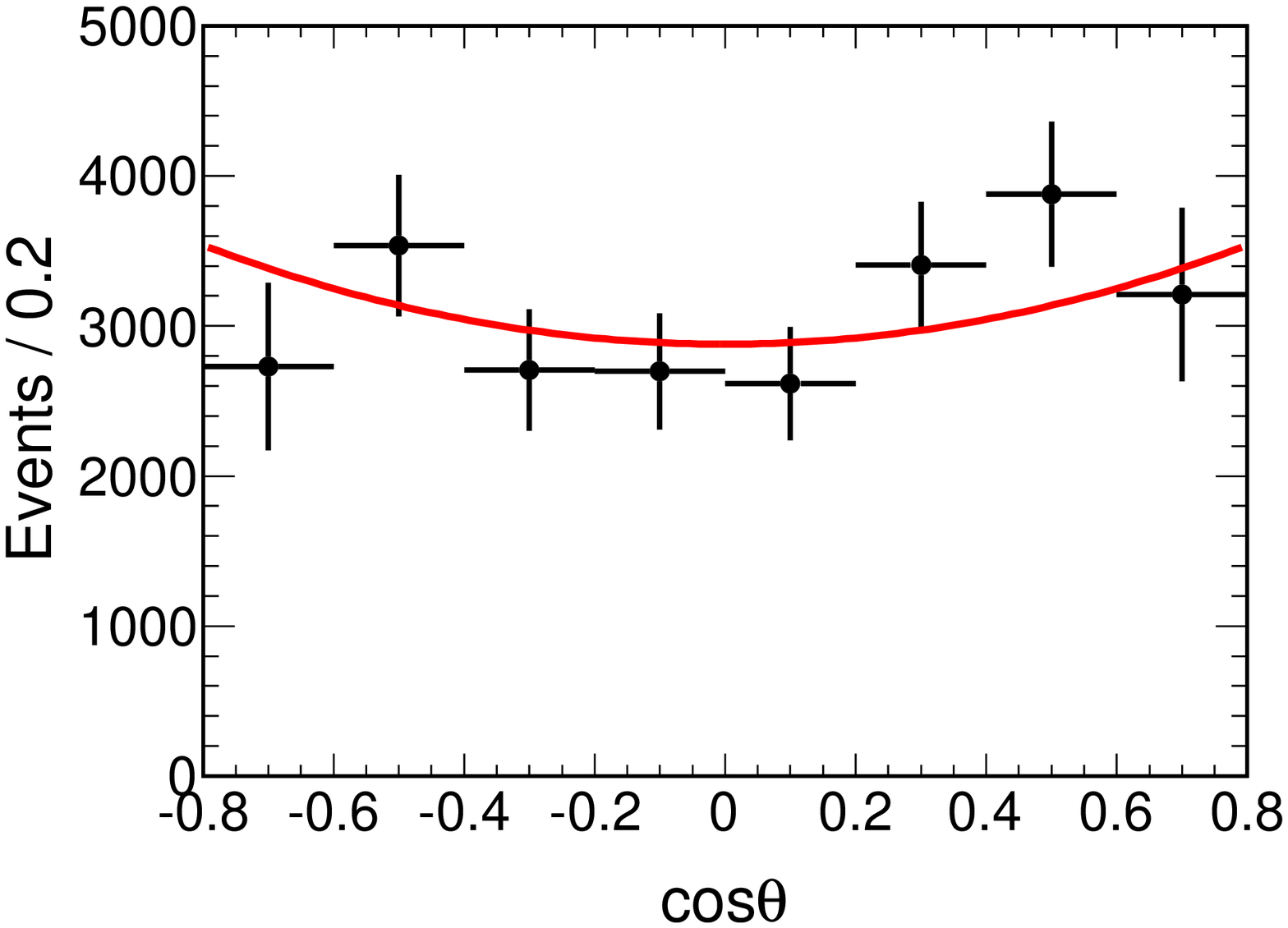}
\caption{Distributions of $\cos\theta_{B}$ for the $\Xi(1530)^{-}$ tag (Left) and the $\bar\Xi(1530)^{+}$ tag (Right). The dots with error bars indicate the efficiency corrected data, and the curves show the fit results.}
\label{angular}
\ecl
\end{figure}

The branching fraction is calculated as 
\begin{equation}\label{BR_equation}
{\cal B}[\psp\ar X]=\frac{N_{\rm obs}}{N_{\psp} \cdot \sum_{i} \epsilon_{i} {\cal B}_{i}},
\end{equation}
where $X$ stands for $\XXXB$ or $\XXBB$,  $N_{\rm obs}$ is the number of extracted signal events,
$N_{\psp}$ is the total number of $\psp$ events~\cite{Ablikim:2017wyh},
$i$ runs over the  $\pi^{-}\Xi^{0}$ and $\pi^{0}\Xi^{-}$ modes,
$\epsilon_{i}$ denotes the detection efficiency obtained with the measured $\alpha$ value for both modes, 
${\cal B}_{i}$ denotes the product of branching fractions of $\Xi(1530)\ar\pi\Xi$ and $\Xi\ar\pi\Lambda$. Table~\ref{result} summarizes the numerical results for the various modes studied. 
The angular distribution parameter for $\psp\ar\XXXB$ decay is determined by performing a least squares fit to the $\cos\theta_{B}$ distribution in the range from $-0.8$ to $0.8$ by Eq.~(\ref{alpha}), divided into 8 equidistant intervals; this is done separately for $\Xi(1530)^-$ and $\bar{\Xi}(1530)^+$  tags.  
The signal yield in each $\cos\theta_{B}$ bin is obtained with the
aforementioned fit method in the $M^{\rm recoil}_{\pi\Xi}$ range of 1.4 GeV/$c^{2}$ to 1.7 GeV/$c^{2}$.
The distributions of the efficiency-corrected signal yields together with the fit curves are
shown in Fig.~\ref{angular}.
The $\alpha$ values obtained 
are summarized in Table~\ref{result}.

\btbl
  \begin{small}
 \caption{The number of the extracted events ($N_{\rm obs}$), efficiencies ($\epsilon_{1}$ is for $\pi^{-}\Xi^{0}$ mode, $\epsilon_{2}$ is for $\pi^{0}\Xi^{-}$ mode),  statistical significance ($S$),  the angular distribution parameter ($\alpha$) and branching fractions (${\cal B}$), where ${\cal B}^{\rm com}$ and  $\alpha^{\rm com}$ denote the combined branching fraction and angular distribution parameters. The first uncertainties are statistical, and the second systematic.}
  \newcommand{\ST}{\rule[0ex]{0pt}{3ex}}
\begin{tabular}{ccccc}  \hline 
\multicolumn{1}{c}{} &\multicolumn{2}{c}{$\psp\ar\XXXB$} &\multicolumn{2}{c}{$\psp\ar\XXBB$}  \\ \hline
Tag mode                &$\Xi(1530)^{-}$ &$\bar\Xi(1530)^{+}$&$\Xi(1530)^{-}$ &$\bar\Xi(1530)^{+}$  \\ \hline
$N_{\rm obs}$              &2664 $\pm$ 114   &2403 $\pm$  132 &152 $\pm$ 37 &247 $\pm$  48\\
$\epsilon_{1}$(\%) &7.85 $\pm$ 0.09  &7.16 $\pm$ 0.08 &8.89 $\pm$ 0.09 &8.42 $\pm$ 0.09 \\
$\epsilon_{2}$(\%) &8.91 $\pm$ 0.09  &8.17 $\pm$ 0.09 &10.58 $\pm$ 0.10 &9.82 $\pm$ 0.10 \\
$S(\sigma)$            &23.0 &18.2 &4.4 &5.3  \\
$\alpha$ &$0.43 \pm 0.30 \pm 0.09$ &$0.36 \pm 0.35 \pm0.08$ &...&...  \\ 
${\cal B}(10^{-5})$ &11.51 $\pm$ 0.49 $\pm$ 0.92 &11.36 $\pm$ 0.62 $\pm$ 1.14 &0.57 $\pm$ 0.14 $\pm$ 0.05 &0.93 $\pm$ 0.18 $\pm$ 0.10 \\
\multicolumn{1}{c}{$\alpha^{\rm com}$} &\multicolumn{2}{c}{$0.40 \pm 0.24 \pm 0.06$} &\multicolumn{2}{c}{...}  \\ 
\multicolumn{1}{c}{${\cal B}^{\rm com}(10^{-5})$} &\multicolumn{2}{c}{11.45 $\pm$ 0.40 $\pm$ 0.59} &\multicolumn{2}{c}{0.70 $\pm$ 0.11 $\pm$ 0.04}  \\ 
\hline
\end{tabular}
\label{result}
 \end{small}
\etbl

Systematic uncertainties on the branching fractions measurements are mainly 
due to differences of detection efficiency between data and 
MC simulation.  
The uncertainties associated with the efficiencies of tracking and PID for the pion from the mother particle $\Xi(1530)^-$ in the $\pi^{-}\Xi^{0}$ decay mode, are investigated with the control sample $\jpsi\ar p\bar{p}\pi^{+}\pi^{-}$. 
The uncertainty due to the 1C kinematic fit for the $\pi^{0}$ reconstruction is estimated with the control sample $\jpsi\ar\rho\pi$.
The uncertainties related to the $\Xi^{0}$ and $\Xi^{-}$ reconstruction efficiency combined with tracking, PID, and the $\Lambda$ reconstruction efficiencies are estimated using the control sample $\psp\ar\XXN$ and $\XXB$.  A detailed description of our methods can be found
in Ref.~\cite{wangxf,Ablikim:2016iym}. 
The uncertainties due to the requirements for mass window and decay length of $\Xi$, $\Lambda$ are estimated with the control sample $\jpsi\ar\XXN$ and $\XXB$. 
The uncertainty related to the mass window of $\Xi(1530)^{-}$ is estimated by varying the half-width of 15 MeV/$c^{2}$ by $\pm 1$ MeV/$c^{2}$.  
The largest difference of the efficiency between data and MC simulation is taken as the systematic uncertainty.
 The uncertainty due to the signal shape is estimated by changing the nominal signal function to the Breit-Wigner function; the difference of the signal yields is taken as the systematic uncertainty.   
The parameters of the Gaussian signal function for the $\XXBB$ final state are fixed in the fit; uncertainties are estimated  by varying the nominal values by 1$\sigma$.
The uncertainty due to the  $M^{\rm recoil}_{\pi\Xi}$ fitting range is estimated by varying the mass range by $\pm$10 MeV/$c^{2}$. The uncertainties due to the assumed polynomial background shape are estimated by alternate fits using a second or a fourth-order Chebychev function. 
 The uncertainty due to the WCB is estimated by comparing the signal yields between the fits  with and without the corresponding component included in the fit. 
 The uncertainty related with the detection efficiency due to the modeling of the angular distribution of the baryon pairs, represented by the parameter $\alpha$, is estimated for the $\XXXB$ mode by varying the measured $\alpha$ values by 1$\sigma$ in the MC simulation.  
For the $\XXBB$ mode, $\alpha$ is set to zero.  
The uncertainties due to the branching fractions of the intermediate states,  $\Xi\ar\pi\Lambda$ and $\Lambda\ar p\pi$ are  taken to be $0.1$\% and 0.8\% according to the PDG~\cite{PDG2016}. The uncertainty of the branching fraction of $\Xi(1530)\ar\pi\Xi$ is taken conservatively according to the branching fraction of $\Xi(1530)\ar\gamma\Xi$, 4.0\% from the PDG~\cite{PDG2016}. 
The uncertainties due to the total number of $\psp$ events (N$_{\psp}$) are determined with inclusive hadronic $\psp$ decays~\cite{Ablikim:2017wyh}.
The various systematic uncertainties on the branching fraction measurements
 are summarized in Table~\ref{error}. The
total systematic uncertainty is obtained by summing the individual
contributions in quadrature.
\begin{table}[!htbp]
 \begin{small}
\caption{Systematic uncertainties (in \%) and their sources for each measured decay mode.}
    \linespread{1.5}
\scalebox{0.9}[0.9]{
\begin{tabular}{lcccc}  \hline
\multicolumn{1}{l}{} &\multicolumn{2}{c}{$\XXXB$} &\multicolumn{2}{c}{$\XXBB$}  \\ \hline
Source                &$\Xi(1530)^{-}$  &$\bar\Xi(1530)^{+}$ &$\Xi(1530)^{-}$  &$\bar\Xi(1530)^{+}$  \\ \hline
Tracking for pion                          &1.0 &1.0 &1.0 &1.0\\
PID for pion                               &1.0 &1.0 &1.0 &1.0\\
$\pi^{0}$ reconstruction                   &1.0 &1.0 &1.0 &1.0\\
$\Xi^{0}$ reconstruction                   &2.6 &4.8 &2.6 &4.8\\
$\Xi^{-}$ reconstruction                   &3.0 &6.0 &3.0 &6.0\\
Mass window of $\Lambda$                   &0.1 &0.3 &0.1 &0.3\\
Decay length of $\Lambda$                  &0.2 &0.1 &0.2 &0.1\\
$\Xi^{0}$ mass window                      &1.4 &1.8 &1.4 &1.8\\
$\Xi^{-}$  mass window                     &1.0 &1.0 &1.0 &1.0\\
$\Xi(1530)^{-}$ mass window                   &1.0 &1.0 &1.0 &1.0\\
Signal shape                               &2.0 &2.2 &1.0 &1.0\\
Parameterization                           &1.0 &1.0 &... &...\\
Fitting range                              &1.8 &2.0 &5.0 &4.9\\
Background shape                           &1.7 &1.3 &4.5 &2.0\\
Wrong combinations                         &3.3 &2.0 &1.0 &1.0\\
Angular distribution                       &1.0 &1.0 &2.3 &2.5\\
 ${\cal{B}}(\Lambda\ar p\pi)$              &0.8 &0.8 &0.8 &0.8\\
${\cal{B}}(\Xi(1530)\ar\pi\Xi)$            &4.0 &4.0 &4.0 &4.0\\
N$(\psp)$                                  &0.7 &0.7 &0.7 &0.7\\ \hline
Total                                      &8.0 &10.0&9.6 &11.1\\ \hline  
\end{tabular}}
\label{error}
\end{small}
\end{table}

Systematic issues for the measurement of the $\alpha$ include the 
determinations of signal yields in $\cos\theta_{B}$ intervals and 
the $\cos\theta_{B}$ fitting procedure. 
Signal yield systematic uncertainties arise from the fit range, the 
background shape, signal shape and WCB.  
%
These are evaluated with a method similar to the one described above;
the resulting differences with respect to the nominal $\alpha$ values 
are taken systematic uncertainties.
The $\cos\theta_{B}$ fitting uncertainties are estimated by 
re-fitting the $\cos\theta_{B}$ distribution with a different binning and fit 
range.  We divide $\cos\theta_{B}$ into five intervals instead of eight, and 
the change in $\alpha$ is taken as the systematic uncertainty. We also repeat the fit after altering the $\cos\theta_{B}$ range to [$-$0.9, 0.9] or [$-$0.7, 0.7], with the same bin size as the nominal fit. The largest changes of $\alpha$ with respect to the nominal fit are taken as systematic uncertainties.  
All the systematic uncertainties for the $\alpha$ measurement are summarized in Table~\ref{error_ang}, where the total systematic uncertainty is 
the quadratic sum of the contributions.  
\begin{table}[!htbp]
 \begin{small}
\caption{\label{error_ang}Systematic uncertainties (absolute) on the measurement of $\alpha$ value for $\psp\ar\XXXB$ decay.}
    \linespread{1.5}
\begin{tabular}{lcc}  \hline
Source                             &$\Xi(1530)^{-}$ &$\bar\Xi(1530)^{+}$\\ \hline
$M^{\rm recoil}_{\pi\Xi}$ fitting range       &0.05 &0.02\\
Background shape                              &0.03 &0.02\\
Signal shape                                  &0.06 &0.04\\
Wrong combinations                            &0.01 &0.04\\
$\cos\theta_{B}$ binning                      &0.02 &0.01\\
$\cos\theta_{B}$ fitting range                &0.03 &0.04\\ \hline
Total                                         &0.09 &0.08\\ \hline 
\end{tabular}
\end{small}
\end{table}

Combined branching fractions and $\alpha$ values are calculated according to the unconstrained averaging introduced in the PDG~\cite{PDG2016}.
Note that the single-baryon recoil mass method leads to some double-counting of the $\Xi(1530)^{-} \bar\Xi(1530)^{+}$ final-state; MC studies indicate this occurs at a rate of about 10\%.  This is taken into account when combining branching fractions and angular distribution parameters.  
The systematic uncertainties are weighted to properly account for common and uncommon systematic uncertainties using  
$\frac{1}{2}\sum_{i,j(i \neq j)}$
$\frac{\sigma^{\prime}_{i}\sigma_{j}}{\sqrt{
\sigma^{\prime 2}_{i}+ \sigma^{2}_{j}
}}$, where $\sigma$ ($\sigma^{\prime}$) is the systematic uncertainty with (without) common sources, and $i,j$ run over the baryon and anti-baryon tags.  

In summary, using  448.1 million $\psp$ events collected with the BESIII detector at the BEPCII,
we present the observation of  $\psp\ar\XXXB$ and $\XXBB$ decays with the statistical significances of more than 10$\sigma$  and 5.0$\sigma$, respectively, based on a single baryon tag strategy.
The branching fractions for $\psp\ar\XXXB$  and $\XXBB$ are measured to be
(11.45 $\pm$ 0.40 $\pm$ 0.59) $\times$ $10^{-5}$ and (0.70 $\pm$ 0.11 $\pm$ 0.04) $\times$ $10^{-5}$, where the first (second) uncertainty is statistical (systematic). 
The corresponding results are summarized in Table~\ref{result}.  
The observation of the decay $\psp\ar\XXBB$ indicates that the SU(3) flavor symmetry is still broken in the $\psp$ case, which further validates the generality of SU(3) flavor symmetry breaking.  
The measured angular distribution parameter $\alpha$ for $\psp\ar\XXXB$ decay agrees with the theoretical prediction~\cite{ppbref02, ppbref01} with our current errors.  
This offers support, within our limited statistics, for these models 
which include quark mass and electromagnetic effects.  

\section{Acknowledgement}
\label{sec:acknowledgement}
The BESIII collaboration thanks the staff of BEPCII and the IHEP computing center for their strong support. This work is supported in part by National Key Basic Research Program of China under Contract No. 2015CB856700; 
Postdoctoral Natural Science Foundation of China under Contract Nos.   2018M630206, 2017M622347;
National Natural Science Foundation of China (NSFC) under Contracts Nos. 11335008, 11425524, 11521505, 11605042, 11625523, 11635010, 11675184, 11705209, 11735014, 11875115, 11905236; 
Chinese Academy of Science Focused Science Grant; National 1000 Talents Program of China;
the Chinese Academy of Sciences (CAS) Large-Scale Scientific Facility Program; the CAS Center for Excellence in Particle Physics (CCEPP); Joint Large-Scale Scientific Facility Funds of the NSFC and CAS under Contracts Nos. U1532257, U1532258, U1732263; 
CAS Key Research Program of Frontier Sciences under Contracts Nos. QYZDJ-SSW-SLH003, QYZDJ-SSW-SLH040; 100 Talents Program of CAS; INPAC and Shanghai Key Laboratory for Particle Physics and Cosmology; German Research Foundation DFG under Contract No. Collaborative Research Center CRC 1044, FOR 2359; Istituto Nazionale di Fisica Nucleare, Italy; Koninklijke Nederlandse Akademie van Wetenschappen (KNAW) under Contract No. 530-4CDP03; Ministry of Development of Turkey under Contract No. DPT2006K-120470; National Science and Technology fund; The Swedish Research Council; U. S. Department of Energy under Contracts Nos. DE-FG02-05ER41374, DE-SC-0010118, DE-SC-0012069; University of Groningen (RuG) and the Helmholtzzentrum fuer Schwerionenforschung GmbH (GSI), Darmstadt;
Post-doctoral research start-up fees of Henan Province under Contract No. 2017SBH005; Ph.D research start-up fees of Henan Normal University under Contract No. qd16164; Program for Innovative Research Team in University of Henan Province (Grant No.19IRTSTHN018).

\end{document}